# PREDICTIVE FACTORS ASSOCIATED WITH SURVIVAL RATE OF CERVICAL CANCER PATIENTS IN BRUNEI DARUSSALAM.

**MADLI F[1], LEONG E[1], ONG SK[2], LIM E[3], TENGAH KA[4].**
**[1]Faculty of Science, Universiti Brunei Darussalam, Jln Tungku Link, [2]NCD Prevention and Control Unit, Ministry of Health, Commonwealth Drive, [3]Department of Histopathology, Raja Isteri Pengiran Anak Saleha Hospital, and [4]Sultan Hassanal Bolkiah Institute of Education, Universiti Brunei Darussalam, Jln Tungku Link, Brunei Darussalam.**

## ABSTRACT

**Introduction**: Cervical cancer is the third most prevalent cancer among women in Brunei Darussalam. This study aims to report the overall survival rates and associated factors of patients diagnosed with malignant cervical cancer in Brunei Darussalam. **Methods**: A retrospective study of patients diagnosed with cervical cancer from 2007 to 2017 in Brunei Darussalam. The data were obtained from the population-based cancer registry in Brunei Darussalam. Kaplan-Meier survival analysis was used to estimate the overall survival rates at 1-, 3- and 5-year intervals while the log-rank test was used to assess differences in survival between groups. Cox Proportional Hazard (PH) regression analysis was used to examine the association of demographic and clinical factors on the survival of cervical cancer patients. **Results**: A total of 329 registered malignant cervical cancer cases were analyzed. The mean age at diagnosis of patients with cervical cancer was 46.7 ± 12.2 years. There were 28.6% deaths and the overall survival rates at 1, 3 and 5 years were 85.4%, 72.6% and 68.6% respectively. Age at diagnosis, cancer stage and histology types were significant predictive factors for overall survival of the patients diagnosed with cervical cancers when analysed on both log rank tests and Cox PH model. **Conclusion:** Age at diagnosis, cancer stage and histology types were significantly associated with the overall survival rates of cervical cancer patients in Brunei Darussalam. Early detection and management of cervical cancer at early stages should be prioritized to improve the survival rate and quality of cancer care.

**Keywords**:  **Cervical cancer, survival rates, population based cancer registry, Brunei Darussalam, Cox proportional hazard regression.**

## INTRODUCTION

Worldwide, cervical cancer is the fourth most common cancer and the fourth leading cause of cancer death in women, with an estimated of 569,847 new cases and 311,365 death for cervical cancer in 2018.[1,2] It is also the leading cause of cancer death in 42 countries, the vast majority of which are in Sub-Saharan Africa and South Eastern Asia.[2] The incidence and mortality rates of cervical cancer are highest in Africa particularly in Southern Africa, Eastern Africa, and Western Africa. The rates are estimated to be 7 to 10 times lower in North America, Australia/New Zealand, and Western Asia.[1] It was reported that cervical cancer is the third most common disease among women in Asia.[3]

**Corresponding author:** Dr. Elvynna Leong, PhD, Department of Mathematical and Computer Science, Faculty of Science, Jalan Tungku Link, Gadong, Universiti Brunei Darussalam, Brunei Darussalam, BE 1410.
Phone no: +673 2460922 ext 1305; Fax no: +673 2461502
Email: elvynna.leong@ubd.edu.bn

In Brunei Darussalam, the leading cause of deaths is cancer followed by heart diseases. In 2017, cancer, heart diseases, diabetes mellitus and cerebrovascular diseases accounted for 52.5% of the total deaths in Brunei Darussalam and cancer represents 19.3% of the total mortalities in the country.[4] Among Bruneian women, cervical cancer has the third highest incidence and number of deaths.[5] Brunei Darussalam is a developing Southeast Asia nation with an estimated population of 422 000 people made up of multi-ethnic groups consisting of Malays (65.7%), followed by Chinese (10.3%) and other ethnicity (24.0%), majority of which 69.4% reside in the Brunei-Muara district, 16.5% in the Belait district, 11.6% in the Tutong district and remaining 2.5% in the Temburong district.[6]

There are numerous reports in published literatures overseas on the survival rates of patients diagnosed with cervical cancers, however little or no data has been published on the survival rate in Brunei Darussalam. Survival information provides an indication of the performance of cervical cancer control in the healthcare services. Hence, this study was conducted to estimate the overall survival rates and the association of selected demographic and clinical factors in patients diagnosed with cervical cancer in Brunei Darussalam.

## METHODS

This retrospective study included all malignant cervical cancer cases diagnosed from 1st January 2007 to 31st December 2017. De-identified data of cervical cancer patients were provided to the researchers in Excel proforma obtained from Brunei Darussalam Cancer Registry, which included both demographic and clinic-pathological information. Brunei citizens and permanent residents registered with the local health services including those who received cancer diagnosis and treatment from overseas were included in the study. Temporary residents diagnosed with cervical cancer were excluded from the study.

The covariates included for analysis were age at diagnosis (<40, 40-49, 50-59, 60-69, ≥70), district of residence (Brunei-Muara, Tutong, Belait, Temburong), ethnicity (Malay, Chinese, Others), histology types (squamous cell carcinoma (SCC), adenocarcinoma (AC), adenosquamous carcinoma (ASC), others), and cancer stage (localized, regional and distant metastasis). In this study, cancer/tumor stages were grouped using a staging system developed by the SEER program i.e. localized, regional and distant metastasis.[7] Carcinoma in-situ and benign tumours were excluded from this study.

The analysis was carried out using R statistical software. Descriptive statistics was performed to study the distribution of cervical cancer cases according to the covariates. Overall survival rates were calculated using the Kaplan-Meier survival analysis. Overall survival (observed survival or all-cause survival) was defined as the period of time from diagnosis to death or end of follow-up, with no restriction on the cause of death. Patients who were still alive or lost to follow-up at the end of the study period were right-censored. Univariate analysis using log-rank test was used to find differences in survival between groups according to the covariates. Hazard ratios (HR) and 95% confidence intervals (CI) were estimated using multivariate analysis, the Cox Proportional Hazard (PH) model. Similar model was also used to select the significant predictors for cervical cancer patients' survival. For all analyses, the level of statistical significance was set at $p < 0.05$. The Cox PH model was assessed for proportionality assumption and goodness-of-fit tests were performed to confirm the selected model is a good fit model.

This study was approved by the PAP-

RSB Institute of Health Science Research and Ethics Committee (IHSREC) and the Medical and Health Research Ethics committee of Ministry of Health (MHREC), Brunei Darussalam [Ref: UBD/PAPRSBIHSREC/2018/149, dated 21st January 2019].

## RESULTS

A total of 359 malignant cervical cancer cases were collected by the Brunei Darussalam cancer registry from 1st January 2007 to 31st December 2017. A total of 329 cases were used in the analysis after removing cases with incomplete information as well as benign and in-situ cases. Of the 329 cases, 94 (28.6%) deaths were recorded and 235 (71.4%) were still alive at time of follow-up.

The mean age at diagnosis for cervical cancer was 46.7 ± 12.2 years. Table 1 summarises the distribution of cervical cases according to patients' characteristics. More than half (64.4%) of the cervical cancer pa-tients were aged 49 years and younger and only 4.3% were in the age group 70 years and above. Majority of the cervical cancer patients lived in Brunei Muara district (69.8%), followed by Belait (14.9%), Tutong (13.4%) and Temburong (1.9%), which is similar to the population distribution in the four districts.[6] A total of 75.7% of the patients were of Malay ethnicity, 13.7% were of Chinese ethnicity and 10.6% of other ethnicities. This study found that 44.1% of the patients' tumours were diagnosed at regional stage while only 12.5% were in distant metastasis stage. Majority of the patients were diagnosed with histological type SCC (64.4%) followed by AC (19.2%), other histological type (10.6%), and ASC (5.8%).

The survival rates according to the socio-demographic and clinico-pathological characteristics of cervical cancer patients were summarized in Table I. The overall survival rates at 1, 3 and 5 years for cervical cancer patients in Brunei Darussalam were 85.4%,

**Table I: Distribution of cervical cancer cases and 5-year survival rates by selected demographic and clinic-pathological characteristics.**

| Variables | | No. of cases (%)¥ | 5-year SR (%) | Log-rank test |
|---|---|---|---|---|
| **Age at Diagnosis** | < 40 | 101 (30.7) | 71.2 | |
| | 40 – 49 | 111 (33.7) | 81.4 | |
| | 50 – 59 | 65 (19.8) | 68.7 | **p < 0.0001*** |
| | 60 – 69 | 38 (11.5) | 40.3 | |
| | ≥70 | 14 (4.3) | 30.0 | |
| **District of Residence** | Brunei Muara | 220 (69.8) | 69.7 | |
| | Tutong | 42 (13.4) | 60.8 | |
| | Belait | 47 (14.9) | 75.1 | *p* = 0.2000 |
| | Temburong | 6 (1.9) | 80.0 | |
| **Ethnicity** | Malay | 249 (75.7) | 65.5 | |
| | Chinese | 45 (13.7) | 84.9 | *p* = 0.0877 |
| | Other | 35 (10.6) | 69.2 | |
| **Cancer Stage** | Localized | 129 (43.4) | 88.2 | |
| | Regional | 131 (44.1) | 64.6 | **p < 0.0001*** |
| | Distant-metastasis | 37 (12.5) | 12.9 | |
| **Histology Types** | SCC | 212 (64.4) | 73.7 | |
| | AC | 63 (19.2) | 61.4 | |
| | ASC | 19 (5.8) | 32.2 | **p = 0.0112*** |
| | Other | 35 (10.6) | 67.5 | |

¥Percentage of cases for each category is based on the total number of cases. *Statistically significant p<0.05. SR = overall survival rate

**Figure 1**

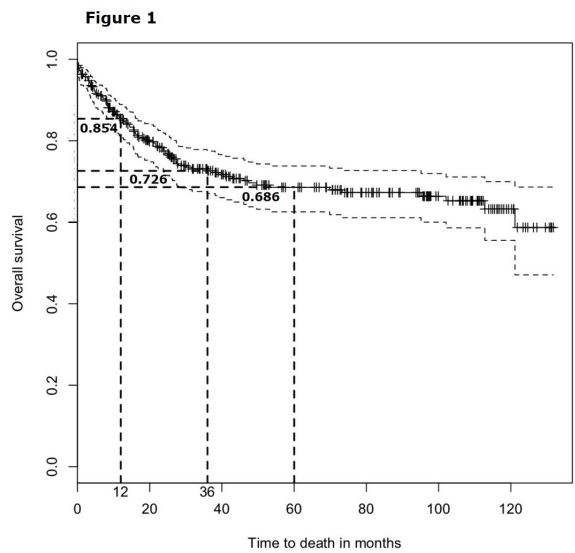

**Figure 2**

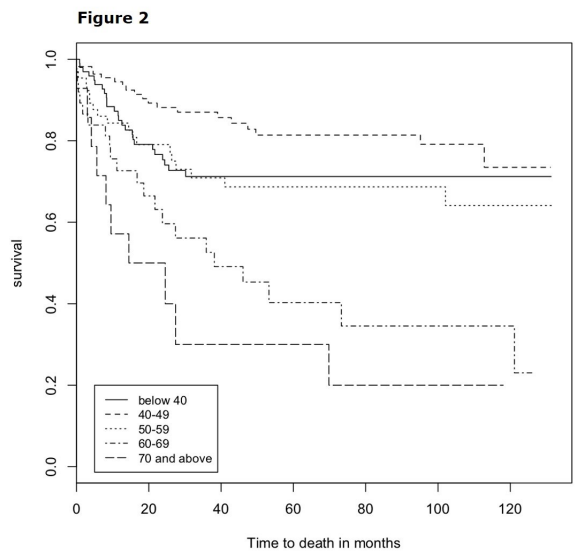

Figure 1: Kaplan Meier Overall Survival Estimates among cervical cancer patients (with 95% C.I). Figure 2: Survival curves by age groups of cervical cancer patients

72.6% and 68.6% respectively as shown in Figure 1. There were survival differences between age groups with those younger than 60 years old achieving overall survival at 5 years of more than 68% compared to those 60 years and older where overall survival dropped below 40% (Figure 2). Racial differences were also noted with Chinese women achieving the highest 5-year survival rate compared to Malay women and other ethnicities although this did not achieved statistical significance (Table I: 84.9% vs 65.5% vs 69.2% respectively; *p=0.0877*). Figure 3 shows patients diagnosed with distant metastasis have the lowest 5-year survival compared to those with localized and regional stages (Table I: 12.9% vs 88.2% vs 64.6% respectively; *p<0.0001*). Patients diagnosed with SCC types had the highest 5-year survival (73.7%) followed by those with AC types (61.4%), other types (67.5%) while patients with ASC type had the lowest survival (32.2%) as shown in Figure 4 (Table I: *p=0.0112*).

The log-rank tests showed that there were significant differences in survival rates between age at diagnosis (Table I: *p <*

*0.0001*), histology types (*p = 0.0112*) and cancer stage (*p < 0.0001*). The results from the Cox PH regression model in Table II further confirmed that all three covariates, age at diagnosis, cancer stage and histology types were statistically significant independent predictive factors for overall survival in cervical cancer patients. No violations of proportional hazard assumptions were observed. For age at diagnosis, the hazards ratio for death showed a bimodal distribution with increased hazards for death in those below 40 years of age [(HR = 2.398; 95% C.I. = (1.259, 4.568); *p=0.0078*] and for those more than 60 years of age [HR(60-69) = 2.894; 95% C.I. = (1.448, 5.786); *p=0.0026*; HR(≥70) = 5.559; 95% C.I. = (2.353, 13.1330; *p<0.0001*].

For cancer stages, compared to localized stage, the expected hazards for death were 3.61 times higher for a cervical cancer patient at regional stage [HR = 3.61, 95% C.I. = (1.89 - 6.89); *p=0.0001*] and 16.76 times higher for a cervical cancer patient with distant metastasis stage [HR = 16.76, 95% C.I. = (8.31 - 33.79); *p<0.0001*]. For histology types, the expected hazard ratio for death for a cervical cancer patient with ASC type was 2.76 times higher as compared to a cervical cancer patient with SCC type [HR = 2.76, 95% C.I. = (1.38 - 5.53); *p=0.0042*].

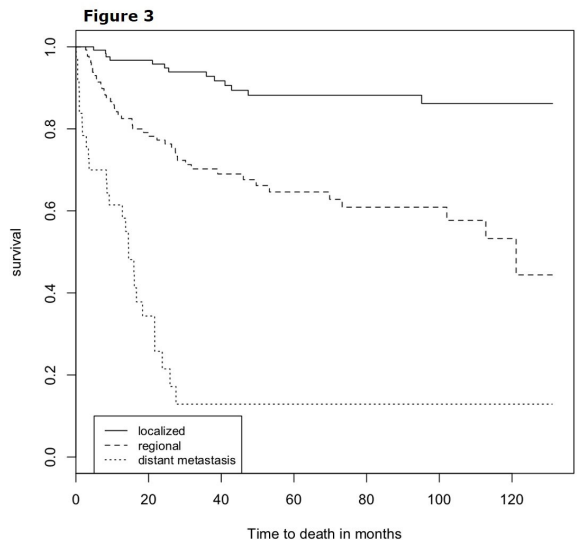
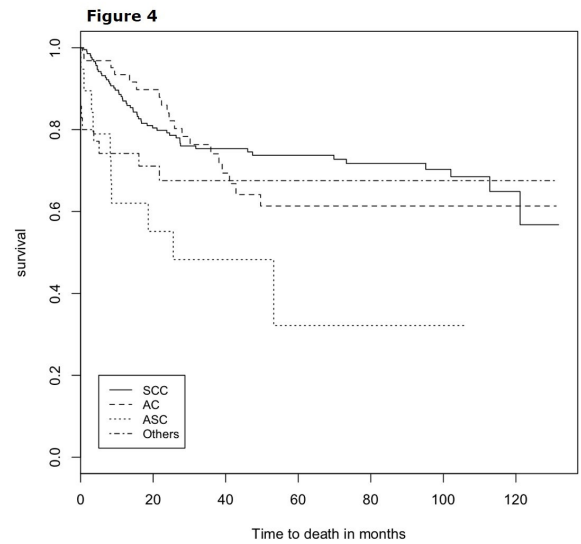

**Figure 3:** Survival curves by cancer stage of cervical cancer patients. **Figure 4:** Survival curves by morphology (histology) of cervical cancer patients.

## DISCUSSION

This is the first reported study on the overall survival rate of patients diagnosed with cervical cancer in Brunei Darussalam. Survival studies provide quantitative information on both survival and factors associated with the survival of patients with a particular disease of interest in a specific region or country. The information will be useful to policy makers and programme planners to review and prioritise healthcare resources and strategies in order to achieve better clinical and survival outcome.

The 5-year overall survival rate of cervical cancer patients in Brunei Darussalam was 68.6%. It was reported that Singapore and Malaysia have 5-year overall survival rates of 54.0% (2011-2015)[8] and 71.1% (2000-2005)[9] respectively. Developed coun-

**Table II: Multivariate analysis of associated factors for cervical cancer patients' survival**.

| Variables | | HR | 95% C.I. | p-value* |
|---|---|---|---|---|
| **Age at Diagnosis** | 40 – 49 (Reference) | 1.000 | - | **-** |
| | <40 | 2.398 | (1.259, 4.568) | 0.0078 |
| | 50 – 59 | 1.863 | (0.946, 3.668) | 0.0719 |
| | 60 – 69 | 2.894 | (1.448, 5.786) | 0.0026 |
| | ≥70 | 5.559 | (2.353, 13.133) | < 0.0001 |
| **Cancer Stage** | Localized (Reference) | 1.000 | - | - |
| | Regional | 3.606 | (1.889, 6.885) | 0.0001 |
| | Distant-metastasis | 16.758 | (8.313, 33.785) | < 0.0001 |
| **Histology Types** | SCC (Reference) | 1.000 | - | - |
| | AC | 1.583 | (0.876, 2.861) | 0.1286 |
| | ASC | 2.758 | (1.376, 5.528) | 0.0042 |
| | Other | 0.630 | (0.247, 1.609) | 0.3343 |

HR = Hazard Ratio, C.I. = Confidence Interval. *p-value for Wald statistics

tries such as Germany, USA, Korea and Australia have 5-year survival rates of 64.7% (2002-2006)[10], 79.9% (2011-2015)[11] and 72.1% (2009-2013)[12] respectively. However, direct comparison should be made with caution between countries due to differences in study time period, socio-demographic factors and healthcare surveillance system.

In this study, histological types has a significant effect on the survival of patients diagnosed with cervical cancer in Brunei Darussalam. This is consistent with other studies which reported that certain histological types were significant prognostic factors of patient survival.[13,14] The 5-year overall survival rates of histological types SCC and AC were 73.7% and 61.4% respectively, which suggests that the survival in women with SCC is better than those with AC. Several other studies have also reported similar survival rates with SCC histological types of cervical cancer compared to AC.[15-17]

This study also confirmed that cancer stage is associated with survival of cervical cancer patients. This study showed that 43.4% cervical cancer cases were at localized stage, 44.1% at regional and 12.5% at distant stage, with the latter stage having the worst 5-year overall survival rates. Previous studies have also reported tumor stage to be a significant prognostic factor for survival of cervical cancer patients.[14,18]

Several studies suggested that survival is associated with age at diagnosis in which young women were found to have better survival.[13,14,19] Our study found similar result where hazard ratio for death from cervical cancer increases with age from 60 years and above. Two recent studies have reported similar increased mortality rates with age.[20,21] In particular Hammer et al from Denmark reported that older women, especially those aged 65 years and above diagnosed with cervical cancer have higher mortality, as much as five times when compared to those aged 40-45 years.[21] She reported that older women above 65 years of age were often not automatically screened for cervical cancer and were often diagnosed with cervical cancer which were at advanced cancer stage, such that they were no longer resectable. Another interesting finding of our study was that survival rate among women younger than 40 years old was significantly lower compared to those aged 40-49 year old. Lau et al also reported poorer survival among younger women aged 30 years or younger who were diagnosed with cervical cancer in Taiwan.[22] He found higher rates of parametrial involvement and distant metastases in this younger group of women suggesting an aggressive nature of the cervical cancer.

In our study, ethnicity and place of residence (district) did not appear to be significant associated factors for survival of cervical cancer patients. Two previous studies have reported ethnicity as a predictor for survival.[23,24] It is important to consider the confounding or combinative effect of other factors associated with cancer survival rates, these may include access to healthcare services and other comorbidities and risk factors such as smoking.[25,26]

In Brunei Darussalam, an organized national cervical cancer screening programme was rolled out in 2011. National cervical cancer screening programmes has been shown to improve early detection and management of cancer and improve prognosis and survival of patients.[27] Similarly, those who attended screening programmes showed better survival as screening allows early detection and treatment of invasive cervical cancer if followed up by appropriate care.[14,28] In addition, HPV vaccination is provided free of charge to female adolescents in Year 7 of secondary schools since 2012. High-risk types of human papillomavirus (HPV), the most common sexually transmitted infections are responsible for the transmitted infections are responsible for the development of cervical cancers. HPV vaccination is recommended to girls aged between 9 and 13 years old before initiation of sexual activity is an important primary preventive intervention against the development of cervical cancer.[29] Currently, there is no data on the prevalence of HPV genotypes in cervical cancer in Brunei. The vaccines currently approved for use cover two of the high risk genotypes of HPV. Future studies should look into the association of survival rates with HPV genotypes, uptake rate of the cervical screening programme and vaccination. An improvement in cervical cancer survival trend can be expected if effective cervical cancer screening and vaccination programmes are sustained to function as both the primary and secondary prevention strategies in controlling the disease.

This study demonstrates the overall survival rate of cervical cancer patients in Brunei Darussalam and the associated factors despite the presence of a few limitations. One limitation is that information such as socioeconomic status, family history of cervical cancer, HPV vaccination and screening attendance, lifestyle practices such as smoking, diet and physical activity were not available to researchers for this study. Future studies should collect these data to provide further insight on the preventive strategies addressing such risk factors. As with any population based disease registry, there were missing data on some variables. This has been minimized as data documentation has been improved with the implementation of electronic medical record system by the Brunei Darussalam healthcare services in 2013.

## CONCLUSION

The overall survival rates at 1, 3 and 5 years for patients diagnosed with cervical cancer in Brunei Darussalam were 85.4%, 72.6% and 68.6% respectively. This study suggests that age at diagnosis, histological type and cancer stage were significant independent predictive factors for survival of patients diagnosed with cervical cancer in Brunei Darussalam. Early detection and management of cervical cancer at early stages should be prioritized to improve survival rate and quality of cancer care in Brunei Darussalam.

## CONFLICT OF INTEREST

The authors declare that there is no conflict of interest.

**Figure 1**

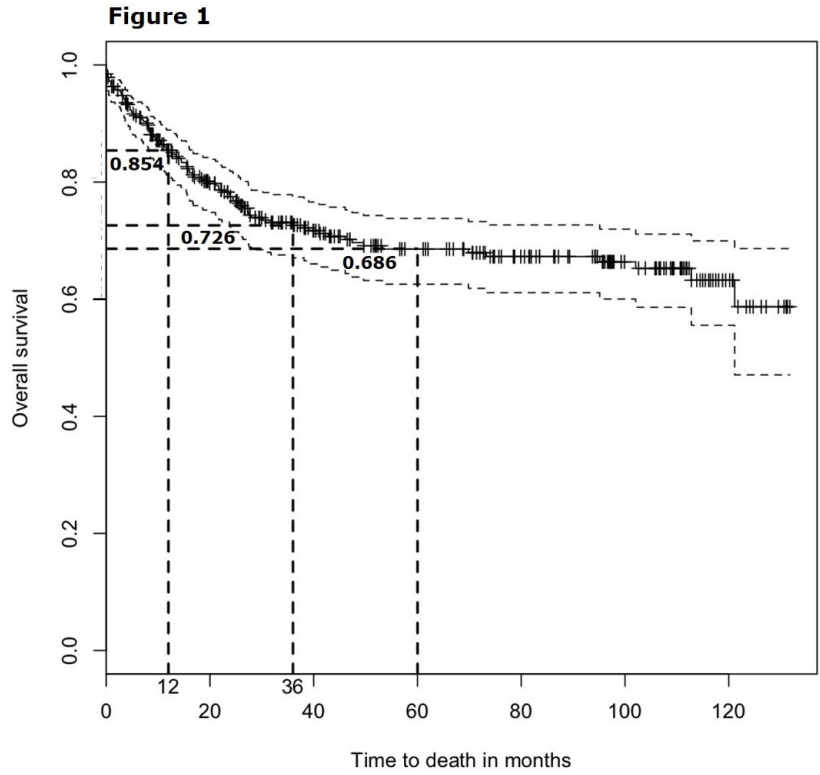

**Figure 2**

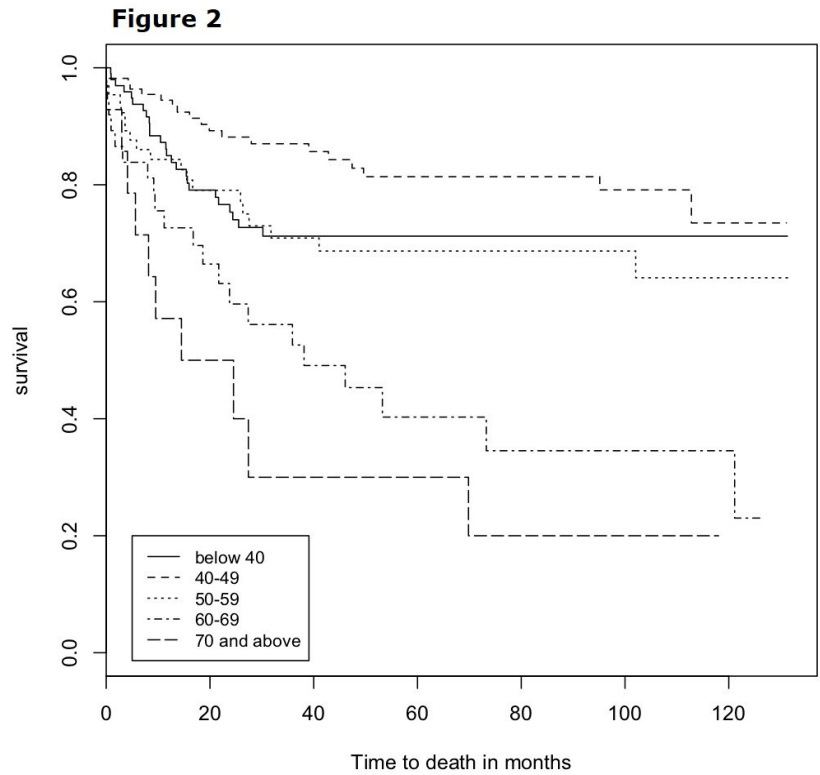

**Figure 3**

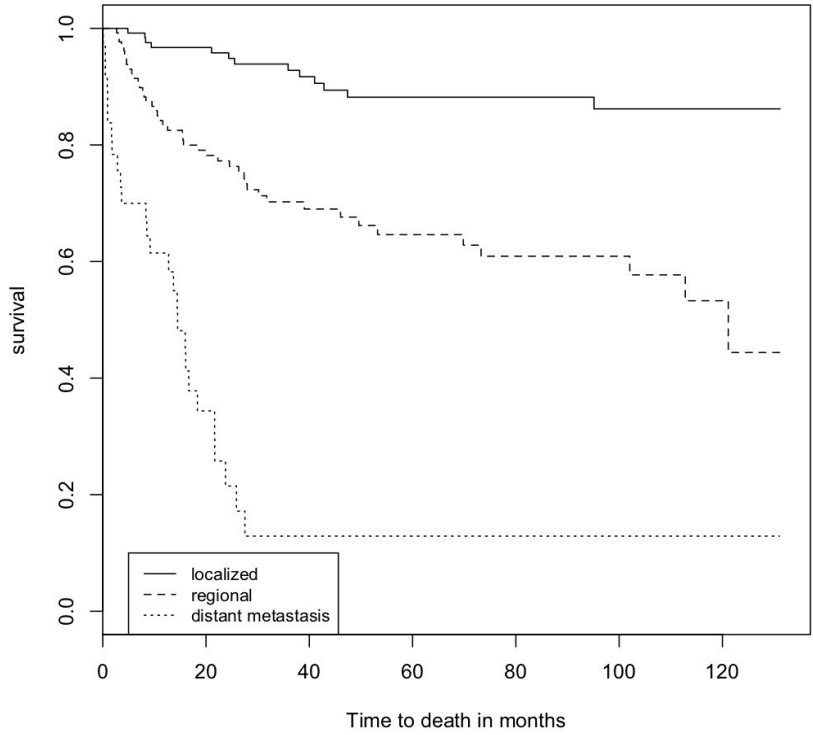

**Figure 4**

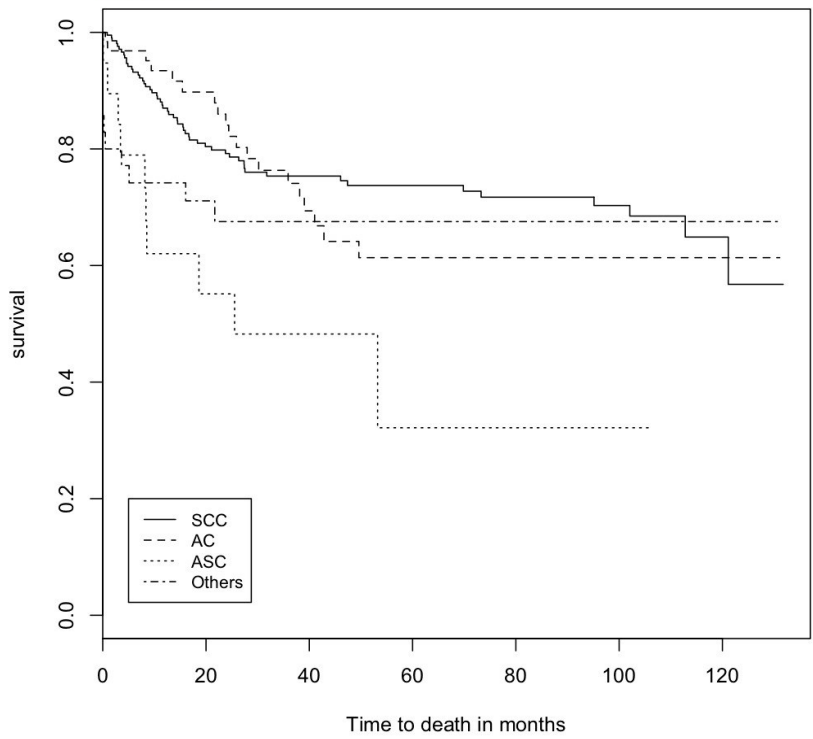